\documentstyle[12pt]{article}

\begin{document}

\raggedbottom
\baselineskip=15pt
\parskip=15pt
\pagestyle{empty}
\begin{flushleft}
{\Large\bf Anisotropic and inhomogeneous cosmologies}
\vspace*{5mm} \\
{\large M.A.H. M{\sc ac}CALLUM}
\vspace*{15pt} \\
School of Mathematical Sciences \\
Queen Mary and Westfield College, University of London \\
Mile End Road, London E1 4NS, U.K. \\
E-mail: M.A.H.MacCallum@qmw.ac.uk
\vspace{48pt} \\
\end{flushleft}

\section{INTRODUCTION}
My first impressions of Dennis Sciama came from a short introductory
astrophysics course he gave to undergraduates in 1964. Then in 1966-7 I
took his Cambridge Part III course in relativity, in which he
charitably ignored my inadvertent use of
Euclidean signature in the examination (an error I spotted just at the
very end of the allowed time) and gave me a good mark. In both these courses
he showed the qualities of enthusiasm and encouragement of students
with which I was to become more familiar later in 1967 when I began as
a research student. A project on stellar structure had
taught me that I did not want to work on that, and I began under Dennis
with the idea of looking at galaxy formation. However, by sharing an
office with John Stewart I came to read John's paper with George Ellis
\cite{SteEll68} and its antecedent \cite{Ell67}
and developed an interest in relativistic cosmological models, which
led to George becoming my second supervisor.

I was still in Sciama's group, and I learnt a lot from the
tea-table conversations, which seemed to cover all of general
relativity and astrophysics. Dennis taught us by example that the
field should not be sub-divided into mathematics and physics, or
cosmological and galactic and stellar, but that one needed to know
about all those things to do really good work. Of course he was not
uncritically enthusiastic: his own opinions were strongly enough held that
we used to joke that if we wanted to stop research all we need do was
say loudly in the tea-room that we did not believe in Mach's
principle.  But it was a very supportive and stimulating atmosphere
for which I will always be grateful.
\parskip = 0pt
\vspace{15pt}

 This is a review of what we have learnt from the study of
non-standard cosmologies in which I got involved 25 years ago. Only
exact solutions will be considered: the perturbation theory will be
left for others to discuss. In
earlier reviews \cite{Mac79,Mac84} I started from
the mathematical classification of the solutions but here I want to
take a different route and consider the application areas. So let me
just quickly remind readers of the general groups of models to which I
will later refer. They are:
 \begin{list}{[\arabic{enumi}]}{\setlength{\itemsep}{0pt}
  \setlength{\partopsep}{0pt}   \setlength{\topsep}{0pt}
  \setlength{\parsep}{0pt}\usecounter{enumi}}
 \item{Spatially-homogeneous and isotropic models. In relativity these give the
Friedman-Lema\^{\i}tre-Robertson-Walker cosmologies, and the
``standard model''.}
 \item{Spatially-homogeneous but anisotropic models. These are the
Bianchi models, in general, the exceptions being the Kantowski-Sachs
models with an $S^2 \times R^2$ topology.}
 \item{Isotropic but inhomogeneous models. These are spherically
symmetric models, whose dust subcases, having been first discussed by
Lema\^{\i}tre (1933)\nocite{Lem33}, are called Tolman-Bondi models.}
 \item{Models with two ignorable coordinates, usually with a pair of
commuting Killing vectors. These may be plane or cylindrically symmetric.}
 \item{Models with less symmetry than those above. Only a few special
cases are known exactly.}
 \end{list}
\vspace{15pt}

In giving this review I only had time to mention and discuss some
selected papers and issues, not survey the whole vast field. Thus the
bibliography is at best a representative selection from many worthy
and interesting papers, and authors whose work is unkindly omitted may
quite reasonably feel it is {\em un}representative.
\vspace{15pt}

\section{OBSERVATIONS, THE STANDARD MODEL AND ALTERNATIVES}\vspace{15pt}

 What is it that a cosmological model should explain? There are
the following main features:
 \begin{list}{[\arabic{enumi}]}{\setlength{\itemsep}{0pt}
  \setlength{\partopsep}{0pt}  \setlength{\topsep}{0pt}
  \setlength{\parsep}{0pt}\usecounter{enumi}}
 \item{Lumpiness, or the clumping of matter. The evidence for this is
obvious.}
 \item{Expansion, shown by the Hubble law.}
 \item{Evolution, shown by the radio source counts and more recently
by galaxy counts.}
 \item{A hot dense phase, to account for the cosmic microwave
background radiation (CMWBR) and the abundances of the chemical elements.}
 \item{Isotropy, shown to a high degree of approximation in various
cosmological observations, but especially in the CMWBR.}
 \item{Possibly, homogeneity. (The doubt indicated here will be
explained later.)}
 \item{The numerical values of parameters of the universe and its
laws, such as the baryon number density, the total density parameter
$\Omega$, the entropy per baryon, and the coupling constants}
 \item{(Perhaps) such features as the presence of life.}
 \end{list}
\vspace{15pt}

The standard big-bang model at the time I started as a student was:
 \begin{list}{[\arabic{enumi}]}{\setlength{\itemsep}{0pt}
  \setlength{\partopsep}{0pt}  \setlength{\topsep}{0pt}
  \setlength{\parsep}{0pt}\usecounter{enumi}}
 \item{Isotropic at all points and thus necessarily\ldots}
 \item{Spatially-homogeneous, implying Robertson-Walker geometry.}
 \item{Satisfied Einstein's field equations}
 \item{At recent times (for about the last $10^{10}$ years)
pressureless and thus governed by the Friedman-Lema\^{\i}tre dynamics.}
 \item{At early times, radiation-dominated, giving the Tolman dynamics
and a thermal history including the usual account of nucleogenesis and
the microwave background.}
 \end{list}
\vspace{15pt}

To this picture, which was the orthodox view from about 1965-80, the
last decade added the following extra orthodoxies:
 \begin{list}{[\arabic{enumi}]}{\setlength{\itemsep}{0pt}
  \setlength{\partopsep}{0pt}  \setlength{\topsep}{0pt}
  \setlength{\parsep}{0pt}\usecounter{enumi}}
 \setcounter{enumi}{5}
 \item{$\Omega = 1$. Thus there is dark matter, for which the Cold
Dark Matter model was preferred.}
 \item{Inflation -- a period in the early universe where some field
effectively mimics a large cosmological constant and so causes a
period of rapid expansion long enough to multiply the initial length
scale many times.}
 \item{Non-linear clustering on galaxy cluster scales, modelled by the
$N$-body simulations which fit correlation functions based on observations.}
 \end{list}
and also added, as alternatives, such concepts as cosmic strings, GUTs or
TOEs\footnote{Why so anatomical?} and so on.
\parskip=15pt

The standard model has some clear successes: it certainly fits the
Hubble law, the source count evolutions (in principle if not in
detail), the cosmic microwave spectrum, the chemical abundances, the
measured isotropies, and the assumption of homogeneity. Perhaps its
greatest success was the prediction that the number of neutrino
species should be 3 and could not be more than 4, a prediction now
fully borne out by the LEP data.

However, the model still has weaknesses \cite{Mac87}.  For
example, the true clumping of matter on large scales, as shown by the
QDOT data \cite{SauFreRow91} and the angular correlation functions of
galaxies \cite{MadEfsSut90}, is too strong for the standard cold dark
matter account\footnote{These discoveries made it possible for
disagreement with the 1980s dogmatism on such matters to at last be
listened to.}. The uniformity of the Hubble flow is under question
from the work of the ``Seven Samurai'' \cite{LynFabBur88} and others. The
question of the true value of $\Omega$ has been re-opened, partly
because theory has shown that inflation does not uniquely predict
$\Omega\:=\:1$ and partly because observations give somewhat variant
values. Some authors have pointed out that our knowledge of the
physics valid at nucleogenesis and before is still somewhat uncertain,
and we should retain some agnosticism towards our account of those
early times.

 Finally, we should recognize that our belief in homogeneity has very
poor observational support. We have data from our past light cone (and
those of earlier human astronomers) and from geological records
\cite{Hoy62}.  Studying homogeneity requires us to know about
conditions at great distances {\em at the present time}, whereas what
we can observe at great distances is what happened a long time ago, so
to test homogeneity we have to understand the evolution both of the
universe's geometry and of its matter content%
\footnote{Local
measures of homogeneity merely tell us that the spatial gradients of
cosmic quantities are not too strong near us.}. Thus we cannot test
homogeneity, only check that it is consistent with the data and our
understanding of the theory.  The general belief in homogeneity is
indeed like the zeal of the convert, since until the 1950s, when Baade
revised the distance scale, the accepted distances and sizes of
galaxies were not consistent with homogeneity.

 These comments, however, are not enough to justify examination of
other models. Why do we do that? I think there are several reasons.
Alternative models provide fully non-linear modelling of local
processes.  They may show whether characteristics thought to be
peculiar to the standard model, and thus a test of it, can occur
elsewhere.  They may be used in attempted proofs that no universe
could be anisotropic or inhomogeneous, by proving that any strong
departures from the standard model decay away during evolution. They
can be comparators in data analysis, to show that only standard models
fit. Finally, they may even be advanced as replacements of the
standard model.

 Before starting to examine how the alternatives fare in these various
r\^{o}les, I must point out two major defects in work up to now. One is
that the matter content is almost always assumed to be a perfect
fluid. Yet even in the simplest non-standard models, the Bianchi
models, as soon as matter is in motion relative to the homogeneous
surfaces (i.e.\ becomes `tilted')\footnote{Such models have recently
been used to fit the observed dipole anisotropy in the CMWBR
\cite{Tur92}, though other explanations seem to me more credible.} it
experiences density gradients which should lead to heat fluxes
\cite{BraSvi84}: similar remarks apply to other simple
models. Attempting to remedy this with some other mathematically
convenient equation of state is not an adequate response; one must try
to base the description of matter on a realistic model of microscopic
physics or thermodynamics, and few have considered such questions
\cite{SalSchLee87,BonCol88,RomPav92}.  The other
objection is that we can only explore the mathematically tractable
subsets of models, which may be far from representative of all models.

\section{MODELLING LOCAL NON-LINEARITIES}
 Cosmic strings have been modelled by cylindrically symmetric models,
starting with the work of Gott \nocite{Got85}, Hiscock
\nocite{His85} and Linet \nocite{Lin85} in 1985. These studies have
usually been done with static strings\footnote{There is some
controversy about whether these can correctly represent strings
embedded in an expanding universe \cite{ClaEllVic90}.}, and have
considered such questions as the effects on classical and quantum
fields in the neighbourhood of the string.

 Similarly, exact solutions for domain walls, using plane symmetric
models, usually static, have been considered \cite{Vil83,IpsSik84,Goe90,Wan91}%
\footnote{Note that since the sources usually have a
boost symmetry in the timelike surface giving the wall, corresponding
solutions have timelike surfaces admitting the
(2+1)-dimensional de Sitter group.}.

 Galactic scale inhomogeneities have frequently been modelled by
spherically symmetric models, usually Tolman-Bondi. They have been
used to study galaxy formation (e.g.\ Tolman (1934), Carr and
Yahil (1990), and Meszaros (1991)\nocite{Tol34,CarYah90,Mes91}),
to estimate departures from the simple theory of the
magnitude-redshift relations based on a smoothed out
model\footnote{The point is that the beams of light we observe are
focussed only by the matter actually inside the beam, not the matter
that would be there in a completely uniform model.} (e.g.\
Dyer (1976), Kantowski (1969b) and Newman (1979)%
\nocite{Dye76,Kan69b,New79}: note that these works show that the
corrections depend on the choice of modelling), and as the simplest
models of gravitational lenses\footnote{The very detailed modern work
interpreting real lenses to study various properties of individual
sources and the cosmos mostly uses linearized approximations.}.
Spherically symmetric inhomogeneities have also been used to model the
formation of primordial black holes \cite{CarHaw74}.

On a larger scale still inhomogeneous spacetimes have been used to
model clusters of galaxies \cite{Kan69b}, variations in the Hubble
flow due to the supercluster \cite{Mav77}, the evolution of cosmic
voids \cite{Sat84,HauOlsRot83,BonCha90}\nocite{BonCha91}, the observed
distribution of galaxies and simple hierarchical
models of the universe \cite{Bon72,Wes78,Wes79,Rib92a}.

The references just cited are only the tip of the iceberg. For his
mammoth survey of all inhomogeneous cosmological models which contain,
as a limiting case, Friedman-Robertson-Walker models, Krasinski now
has read about 1900 papers (as reported at the GR13 conference in 1992%
\footnote{The survey is not yet complete and remains to be published,
but interim reports have appeared in some places, e.g.\ \cite{Kra90}.}).
As well as the issues mentioned above, these papers discuss
many others including models for interactions between different forms
of matter, generation of gravitational radiation, and the nature of cosmic
singularities.

There is not enough space here to do all these arguments justice, and
anyway it would be unfair to pre-empt Krasinski's conclusions.
Moreover, I believe the issues for which I have given a few detailed
references are (together with some appearing later in this survey) the
most important astrophysically. So I will just mention two more points
which have arisen. One is that some exact non-linear solutions obey
exactly the linearized perturbation equations for the FLRW models
\cite{GooWai82,CarChaFei83}. The other is a {\em jeu d'esprit} in
which it was shown that in a ``Swiss cheese'' model, made by joining
two FLRW exteriors at the two sides of a Kruskal diagram for the
Schwarzschild solution, one can have two universes each of which can
receive (but not answer) a signal from the other \cite{Sus85}.

\section{WHICH FLRW PROPERTIES ARE SPECIAL?}
The earliest use of anisotropic cosmological models to
study a real cosmological problem was the investigation by Lema\^{\i}tre
(1933) \nocite{Lem33} of the occurrence of singularities in Bianchi type I
models. The objective was to explore whether the big-bang which arose
in FLRW models was simply a consequence of the assumed symmetry: it
was of course found not to be.

A later similar investigation was to see if the helium abundance, as
known in the 1960s, could be fitted better by anisotropic cosmologies
than by FLRW models, which at the time appeared to give discrepancies.
The reason this might happen is that anisotropy speeds up the
evolution between the time when deuterium can first form, because it
is no longer dissociated by the photons, and the time when neutrons
and protons are sufficiently sparse that they no longer find each
other to combine. Hawking and Tayler (1966) \nocite{HawTay66} were
pioneers in this effort, which continued into the 1980s but suffered
some mutations in its intention.

First the argument was reversed, and the good agreement of FLRW
predictions with data was used to limit the anisotropy during the
nucleogenesis period (see e.g.\ Barrow (1976), Olson (1977)%
\nocite{Bar76,Ols77}). Later still
these limits were relaxed as a result of considering the effects of
anisotropic neutrino distribution functions \cite{RotMat82}
and other effects on reaction rates \cite{JusBajGor83}. It has
even been shown \cite{MatVogMad84,Bar84} that strongly anisotropic
models, not obeying the limits deduced from perturbed FLRW models, can
also produce correct element abundances, though they may violate other
constraints \cite{MatMad85,MatMadVog85}.

The above properties turn out not to be special to FLRW geometry. One
that might be thought to be is the exact isotropy of the CMWBR. To
test this, many people in the 1960s and 70s computed the angular
distribution of the CMWBR temperature in Bianchi models (e.g.\
Thorne (1967), Novikov (1968), Collins and Hawking (1972), and Barrow
{\em et al.\/} (1985)%
\nocite{Tho67,Nov68,ColHaw72,BarJusSon85}). These
calculations allow limits to be put on small deviations from isotropy
from observation, and also enabled, for example, the prediction of `hot
spots' in the CMWBR in certain Bianchi models, which could in principle
be searched for, if there were a quadrupole component\footnote{Which,
since the meeting this survey was given at, has been shown to exist in
the COBE data.}, to see if the quadrupole verifies one of those models.

Similar calculations, by fewer people, considered the polarization
\cite{Ree68,Ani74,TolMat84} and spectrum \cite{Ree68,Ras71}. More
recently still, work has been carried out on the microwave background
in some inhomogeneous models \cite{SaeArn90}. It has been shown that pure
rotation (without shear) is not ruled out by the CMWBR \cite{Obu92},
but this result may be irrelevant to the real universe where shear is
essential to non-trivial perturbations \cite{Goo83,Dun92}.

One property, the nature of the big-bang singularity, as distinct from
its existence, has been so extensively discussed as to demand a
section of its own.

\section{THE ASYMPTOTIC BEHAVIOUR OF CLASSICAL COSMOLOGIES}
One can argue that classical cosmologies are irrelevant before the
Planck time, but until a theory of quantum gravity is established and
experimentally verified (if indeed that will ever be possible) there
will be room for discussions of the behaviour of classical models near
their singularities.

In the late 1950s and early 60s Lifshitz and Khalatnikov and their
collaborators showed (a) that singularities in synchronous coordinates
in inhomogeneous cosmologies were in general `fictitious' and (b) that
a special subclass gave real curvature singularities \cite{LifKha63}.
{}From these facts they (wrongly) inferred that general solutions did
not have singularities. This contradicted the later singularity
theorems (for which see Hawking and Ellis (1973)%
\nocite{HawEll73}), a disagreement
which led to the belief that there were errors in LK's arguments.
They themselves, in collaboration with Belinskii, and independently
Misner, showed that Bianchi IX models gave a more complicated,
oscillatory, behaviour than had been discussed in the earlier work,
and Misner christened this the `Mixmaster' universe after a brand of
food mixer.

The detailed behaviour of the Mixmaster model has been the subject of
still-continuing investigations: some authors argue that the evolution
shows ergodic and chaotic properties, while others have pointed out
that the conclusions depend crucially on the choice of time variable
\cite{Bar82,BurBurEll90,Ber91}. Numerical investigations are
tricky because of the required dynamic range if one is to study an
adequately large time-interval, and the difficulties of integrating
chaotic systems.

The extension of these ideas to the inhomogeneous case, by Belinskii,
Lifshitz and Khalatnikov, has been even more controversial, though
prompting a smaller literature. It was strongly attacked by Barrow and
Tipler (1979) \nocite{BarTip79} on a number of technical grounds, but
one can take the view that these were not as damaging to the case as
Barrow and Tipler suggested \cite{BelLifKha80,Mac82}.
Indeed the `velocity-dominated' class whose singularities are like the
Kasner (vacuum Bianchi) cosmology have been more rigorously
characterized and the results justified \cite{EarLiaSac71,HolJolSma90}.
Sadly this does not settle the more general question, and
attempts to handle the whole argument on a completely rigorous footing%
\footnote{One of them made by Smallwood and myself.} have so far failed.

General results about singularity types have been proved. The `locally
extendible' singularities, in which the region around any geodesic
encountering the singularity can be extended beyond the singular
point, can only exist under strong restrictions \cite{Cla76},
while the `whimper' singularities \cite{KinEll73}, in which
curvature invariants remain bounded while curvature components in some
frames blow up, have been shown to be non-generic and unstable
\cite{Sik78}. Examples of these special cases were found among
Bianchi models, and both homogeneous and inhomogeneous cosmologies
have been used as examples or counter-examples in the debate.

A further stimulus to the study of singularities was provided by
Penrose's conjecture that gravitational entropy should be low at the
start of the universe and this would correspond to a state of small or
zero Weyl tensor \cite{Pen79,Tod92}.

Studies of the behaviour of Bianchi models have been much advanced by
the adoption of methods from the theory of dynamical systems. In the
early 70s this began with the discussion of phase portraits for
special cases \cite{Col71} and was extended in
work in which (a) the phase space was compactified, (b) Lyapunov
functions, driving the system near the boundaries of the phase space,
were found and (c) analyticity together with the behaviour of critical
points and separatrices was used to derive the asymptotic behaviour
\cite{Bog85}. In the last decade these methods have been
coupled with the parametrization of the Bianchi models using
automorphism group variables \cite{ColHaw73,Har79,Jan79,Sik80,RoqEll85,Jak87}.

The automorphism group can be briefly described as follows. Writing
the Bianchi metrics as
 $$
 ds^2 = -dt^2 + g_{\alpha\beta}(t)({e^\alpha}_\mu dx^\mu)({e^\beta}_\nu dx^\nu)
 $$
where the corresponding basis vectors $\{{\bf e}_\alpha\}$ obey
 $$
 [{\bf e}_\alpha,\:{\bf e}_\beta ] = {C^\gamma}_{\alpha\beta}{\bf e}_\gamma
 $$
 in which the C's are the structure constants of the relevant symmetry group,
one uses a transformation
 $$ \hat{{\bf e}}^\alpha = {M^\alpha}_\beta {\bf e}^\beta $$
chosen so that the $\{\hat{{\bf e}}_\alpha\}$ obey the same commutation
relations as the  $\{{\bf e}_\alpha\}$. The matrices $M$ are
time-dependent and can be chosen so that the new metric coefficients
$\hat{g}_{\alpha\beta}$ take some convenient form. The real dynamics
is in these metric coefficients. The idea is present in earlier
treatments which grew from Misner's methods for the Mixmaster case
\cite{RyaShe75} but unfortunately the type IX case was highly
misleading in that for Bianchi IX (and no others except Bianchi I) the
rotation group is an automorphism group.

A long series of papers by Jantzen, Rosquist and collaborators
\cite{Jan84,RosUggJan90} have coupled these
ideas with Hamiltonian treatments in a powerful formalism. Using a
different, and in some respects simpler, set of variables,
Wainwright has also attacked the asymptotics problem \cite{WaiHsu89}:
his variables are well-suited for those questions because
their limiting cases are physical evolutions of simpler models rather
than singular behaviours.

The conclusions of these studies have justified the work of Belinskii
{\em et al.\/} for the homogeneous case (but do not affect the arguments about
the inhomogeneous cases) and have enabled new exact solutions to be
found and some general statements about the occurrence of these
solutions (which in general have self-similarity in time) to be made
\cite{WaiHsu89}, in particular showing their r\^{o}les as
attractors of the dynamical systems.

The other class of models where techniques have improved considerably
are the models with two commuting Killing vectors, even when these
vectors are not hypersurface-orthogonal. Some studies have focussed on
the mathematics, showing how known vacuum solutions can be related by
solution-generating techniques \cite{Kit84}, while others have
concentrated on the physics of the evolution of fluid models (not
obtainable by generating techniques, except in the case of `stiff'
fluid, $p\:=\:\rho$) and interpretative
issues \cite{WaiAnd84,HewWaiGla91}. It
emerges that the models studied are typically Kasner-like near the
singularity (agreeing with the LK arguments), and settle down to
self-similar or spatially homogeneous models with superposed
high-frequency gravitational waves at late times. However, some cases
have asymptotic behaviour near the singularity like plane waves, and
others are non-singular \cite{ChiFerSen91}.
The Penrose conjecture has been particularly developed, using exact
solutions as examples, by Wainwright and Goode, who have given a
precise definition to the notion of an `isotropic singularity'
\cite{GooColWai92,Tod92}.

Many authors have also considered the far future evolution (or, in
closed models, the question of recollapse, whose necessity in Bianchi
IX models lacked a rigorous proof until recently \cite{LinWal91}).
{}From various works \cite{Mac71,ColHaw73,BarTip78}
one finds that the homogeneous but anisotropic
models do not in general settle down to an FLRW-like behaviour but
typically generate shears of the order of 25\% of their expansion rates.

This last touches on an interesting question about our account of the
evolution of the universe: is it structurally stable, or would small
changes in the theory of the model parameters change the behaviour
grossly? Several instances of the latter phenomenon, `fragility', have
recently been explored by Tavakol, in collaboration with Coley, Ellis,
Farina, Van den Bergh and others \cite{ColTav92}.

\section{DO NON-FLRW MODELS BECOME SMOOTH?}
Attempts to smoothe out anisotropies or inhomogeneities by any process
obeying deterministic sets of differential equations satisfying
Lipschitz-type conditions are doomed to fail, as was first pointed out
by Collins and Stewart (1971) \nocite{ColSte71} in the context of viscous
mechanisms. The argument is simply that one can impose any desired
amount of anisotropy or inhomogeneity now and evolve the system
backwards in time to reach initial conditions at some earlier time
whose evolution produces the chosen present-day values. This was one
of the arguments which rebutted Misner's ingenious suggestion that
viscosity in the early universe could explain the present level of
isotropy and homogeneity regardless of the initial conditions.

The same argument also holds for inflationary models. Inflation in
itself, without the use of singular equations or otherwise
indeterminate evolutions, cannot wholly explain present isotropy or
homogeneity, although it may reduce deviations by large factors
\cite{Sir82,Wal83,MosSah86,FutRotMat89}. Objections to some specific
calculations have been given \cite{RotEll86}.  Although one can argue
that anisotropy tends to prolong inflation, this does not remove the
difficulty.

Since 1981 I have been arguing a heretical view about one of the
grounds for inflation, namely the `flatness problem', on the grounds
that the formulation of this problem makes an implicit and unjustified
assumption that the {\em a priori} probabilities of values of $\Omega$ is
spread over some range sufficient to make the observed closeness to 1
implausible. Unless one can justify the {\em a priori} distribution, there
is no implausibility\footnote{One can however argue that only
$\Omega\:=\:1$ is plausible, on the grounds that otherwise the quantum
theory before the Planck time would have to fix a length-scale
parameter much larger than any quantum scale, only the $\Omega\:=\:1$
case being scale-free. I am indebted to Gary Gibbons
for this remark.} \cite{Ell91}.

However, if one accepts there is a flatness problem, then there is
also an isotropy problem, since at least for some probability
distributions on the inhomogeneity and anisotropy the models would
not match observation. Protagonists of inflation cannot have it both
ways. Perhaps, if one does not want to just say ``well, that's how the
universe was born'', one has to explain the observed smoothness by
appeal to the `speculative era', as Salam (1990) called it\nocite{Sal90},
i.e.\ by appeal to one's favourite theory of quantum gravity.

Incidentally, one may note that inflation does not solve
the original form of the `horizon problem', which was to account
completely for
the similarity of points on the last scattering surface governed by
different subsets of the inital data surface. Inflation leads to a
large overlap between these initial data subsets, but not to their exact
coincidence. Thus one still has to assume that
the non-overlap regions are not too different. While this may give a
more plausible model, it does not remove the need for assumptions on
the initial data.

A further interesting application of non-standard models has come in a
recent attempt to answer the question posed by Ellis and Rothman
(unpublished) of how the universe can choose a uniform
reference frame at the exit from inflation when a truly de Sitter
model has no preferred time axis. Anninos {\em et al.\/}  (1991a)
\nocite{AnnMatRot91}
have shown by taking an inflating Bianchi V model that the answer is
that the memory is retained and the universe is never really de Sitter.

Finally, one may comment that if inflation works well at early times,
then inflation actually enhances the chance of an anisotropic model fitting the
data, and that since the property of anisotropy cannot be totally
destroyed in general (because it is coded into geometric invariants
which cannot become zero by any classical evolution) the anisotropy
could reassert itself in the future!

\section{\protect\raggedright
ASTROPHYSICAL AND OBSERVABLE CONSEQUENCES OF NON-STANDARD MODELS}

Galaxy formation in anisotropic models has been studied to see if they
could overcome the well-known difficulties of FLRW models (without
inflation), but with negative results \cite{PerMatShe72}.

As mentioned above solutions with two commuting Killing vectors
provide models for universes with gravitational waves. Aspects of
these models have been considered by several authors, e.g.\
Carr and Verdaguer (1983), Ibanez and Verdaguer (1983),
Feinstein (1988)\nocite{CarVer83,IbaVer83,Fei88}. There are in fact
several mathematically related but physically distinct classes of
solutions of the Einstein equations accessible by generating
techniques: stationary axisymmetric spacetimes, colliding wave
solutions (nicely summarized in Griffiths (1991) and Ferrari (1990)%
\nocite{Gri91,Fer90}), and
cosmological solutions, the differences arising from the timelike or
spacelike nature of the surfaces of symmetry and the nature of the
gradient of the determinant of the metric in those surfaces.

The generating techniques essentially work for forms of matter with
characteristic propagation speed equal to the velocity of light, and
use one or more of a battery of related techniques: B\"acklund
transformation, inverse scattering, soliton solutions and so on. One
interesting question that has arisen from recent work is whether
solitons in relativity do or do not exhibit non-linear interactions:
Boyd {\em et al.\/} (1991), \nocite{BoyCenKla91} in investigations of
solitons in a Bianchi I background, found no non-linearity, while
Belinskii (1991) \nocite{Bel91} has claimed there is a non-linear effect.

Work on the observable consequences of non-standard models has been
done by many, as mentioned above. One intriguing possibility raised by
Ellis {\em et al.} (1978) \nocite{EllMaaNel78} is that the observed
sphere on the last scattering surface could lie on a timelike
(hyper)cylinder of homogeneity in a static spherically symmetric
model. This makes the CMWBR isotropic {\em at all points} not only at
the centre, and although it cannot fit all the other data, the model
shows how careful one must be, in drawing conclusions about the
geometry of the universe from observations, not to assume the result
one wishes to prove.

Recent work by Ribeiro (1992b)\nocite{Rib92b}, in the course of an attempt to
make simple models of fractal cosmologies using Tolman-Bondi metrics,
has reminded us of the need to compare data with relativistic models
not Newtonian approximations. Taking the Einstein-de Sitter model, and
integrating down the geodesics, he plotted the number counts against
luminosity distances. At small distances, where a simple
interpretation would say the result looks like a uniform density, the
graph is irrelevant because the distances are inside the region where
the QDOT survey shows things are lumpy
\cite{SauFreRow91}, while at greater redshifts the universe ceases to have a
simple power-law relation of density and distance. Thus even
Einstein-de Sitter does not look homogeneous!

One must therefore ask in general ``do homogeneous models look
homogeneous?''. Of course, they will if the data is handled with
appropriate relativistic corrections, but to achieve such comparisons
in general requires the integration of the null geodesic equations in
each cosmological model considered, and, as those who have tried it
know, even when solving the field equations is simple, solving the
geodesic equations may not be.

Ultimately we will have to refine our understanding with the help of
numerical simulations which can include fully three-dimensional
variations in the initial data, and some excellent pioneering work has
of course been done, e.g.\ Anninos {\em et al.\/} (1991b)%
\nocite{AnnMatTul91}, but capabilities are
still limited (for example Matzner (1991) \nocite{Mat91} could only
use a space grid of $31^3$ points and 256 time steps). Moreover before
one can rely on numerical simulations one needs to prove some
structural stability results.

\section{IS THE STANDARD MODEL RIGHT?}
While I do not think one can give a definitive answer to this
question, I would personally be very surprised if anisotropic but
homogeneous models turned out to be anything more than useful
examples. However, the status of fully inhomogeneous models is less
clear.

One argument is that while the standard models may be good
approximations at present, they are unstable to perturbations both in
the past and the future. The possible alternative pasts are quite
varied, as shown in section 5, even without considering quantum
gravity. Similarly, as also mentioned in section 5, the universe may
not be isotropic in the far future. Moreover, there is the question of
on what scale, if any, the FLRW model is valid. Its use implies some
averaging, and is certainly not correct on small scales. Is it true on
any scale? If so, on what scales? There may be an upper as well as a
lower bound, since we have no knowledge of
conditions outside our past null cone, where some inflationary
scenarios would predict bubbles of differing FLRW universes, and
perhaps domain walls and so on.

If the universe were FLRW, or very close to that, this means it is in a
region, in the space of all possible models, which almost any
reasonable measure is likely to say has very low probability (though
note the remarks on assignments of probabilities in section 6). One
can only evaluate, and perhaps explain, this feature by considering
non-FLRW models. It is noteworthy that many of the ``problems''
inflation claims to tackle are not problems if the universe simply is
always FLRW. Hence, as already argued above, one has a deep problem in
explaining why the universe is in the unlikely FLRW state if one
accepts the arguments about probabilities current in work on inflation.

Moreover, suppose we speculated that the real universe is significantly
inhomogeneous at the present epoch (at a
level beyond that arising from perturbations in FLRW). What would the
objections be? There are only two relevant pieces of data, as far as I
can see. One is the deep galaxy counts made by the automatic plate
measuring machines, which are claimed to restrict variations to a few
percent, and the other is the isotropy of the CMWBR. Although the
latter is a good test for large lumps in a basically FLRW universe,
one has to question (recalling the results of Ellis {\em et al.\/}%
\nocite{EllMaaNel78})
whether it really implies homogeneity.

There is a theorem by Ehlers, Geren and Sachs (1968)
\nocite{EhlGerSac68} showing that if a congruence of
geodesically-moving observers all observe an isotropic distribution of
collisionless gas the metric must be Robertson-Walker. Treciokas and
Ellis (1971) have investigated the related problem with
collisions\nocite{TreEll71}. Recently Ferrando {\em et al.\/} (1992)
have investigated inhomogeneous models where an isotropic gas
distribution is possible\nocite{FerMorPor92}. These studies throw into
focus a conjecture which is usually assumed, namely that an
approximately isotropic gas distribution, at all points, would imply
an approximately Robertson-Walker metric. (It is this assumption which
underlies some of the arguments used, for example, by Barrow in his
talk at this meeting.)

Whether the standard model is correct or not, I feel confident in
concluding that one of the more outstanding inhomogeneities is the
dedicatee of this piece, Dennis Sciama, and I hope some small part of
his talents has been shown here to have been passed on to me. To show
how it has influenced the subject, I have marked authors cited in the
bibliography below who also appear in the Sciama family tree by an asterisk.

I would like to thank G.F.R. Ellis for comments on the first draft of
this survey.



\end{document}